\documentclass[prl,twocolumn,showpacs]{revtex4}% Physical Review Letters

\usepackage{graphicx}% Include figure files
\usepackage{dcolumn}% Align table columns on decimal point
\usepackage{bm}% bold math

\begin{document}

%\preprint{APS/123-QED}

\title{Spectral Properties near the Mott Transition in the One-Dimensional Hubbard Model}

\author{Masanori Kohno}
\affiliation{WPI Center for Materials Nanoarchitectonics, 
National Institute for Materials Science, Tsukuba 305-0044, Japan}

\date{\today}

\begin{abstract}
Single-particle spectral properties near the Mott transition in the one-dimensional Hubbard model are investigated 
by using the dynamical density-matrix renormalization group method and the Bethe ansatz. 
The pseudogap, hole-pocket behavior, spectral-weight transfer, and upper Hubbard band 
are explained in terms of spinons, holons, antiholons, and doublons. The Mott transition is characterized by the emergence of 
a gapless mode whose dispersion relation extends up to the order of hopping $t$ (spin exchange $J$) 
in the weak (strong) interaction regime caused by infinitesimal doping.
\end{abstract}


\maketitle
%{\it introduction.$-$}
High-$T_c$ cuprate superconductors are obtained by doping Mott insulators that have gapless spin and gapped charge excitations. 
In the small-doping regime, anomalous features, such as pseudogaps, Fermi arcs, hole pockets, and kinks in dispersion relations, 
have been observed \cite{kink,holepocket}. 
It is widely believed that understanding such anomalous electronic properties near the Mott transition will be critical for achieving high-$T_c$ superconductivity. 
\par
The Mott transition can be viewed as charge localization caused by strong Coulomb repulsions; 
note that the nature of this transition contrasts with that of conventional metal-to-band-insulator transitions 
where a band of single-particle states is fully occupied regardless of the interactions. 
Further, because of the strong correlations, it is generally difficult to obtain reliable information and intuitive understanding on the Mott transition. 
However, although some detailed properties might depend on the lattice structure, 
it can be expected that essential features of Mott transitions are generally true for finite dimensions 
and that the generic features can be deduced by using a one-dimensional (1D) Hubbard model. 
In this model, we can interpret excitations relevant to the Mott transition using exact solutions without bias. 
Also, we can investigate the properties of a single metallic phase all the way to the Mott transition point 
without causing any instability to phase separation, superconducting, or magnetic orders, which might otherwise have occurred in higher dimensions. 
\par
In this Letter, we investigate the spectral properties near the Mott transition in a 1D Hubbard chain using unbiased numerical techniques 
and exact solutions, focusing attention on the pseudogap, hole-pocket behavior, upper Hubbard band (UHB), and spectral-weight transfer 
from the UHB to the lower Hubbard band (LHB). 
We also discuss the nature of the Mott transition through comparisons with the Fermi liquid picture and 
recent numerical results on the two-dimensional (2D) Hubbard model \cite{ImadaPRL}. 
\par
{\it Model and method.$-$} 
We consider the 1D repulsive Hubbard model defined by the following Hamiltonian: 
$$
{\cal H}=-t\sum_{i,\sigma}\left(c_{i+1\sigma}^{\dagger}c_{i\sigma}+{\mbox {H.c.}}\right)+U\sum_{i}n_{i\uparrow}n_{i\downarrow}-\mu\sum_{i}n_{i}, 
$$
where $c_{i\sigma}$ and $n_{i\sigma}$ are the annihilation and number operators of an electron at site $i$ with spin $\sigma$, respectively, 
and $n_i=n_{i\uparrow}+n_{i\downarrow}$. The hopping integral $t$ and on-site repulsion $U$ are positive.
The number of sites, electrons, and down spins is denoted by $L$, $N$, and $M$, respectively. The doping concentration is defined as $\delta=1-N/L$. 
Also, we define the single-particle spectral function as follows: 
\begin{eqnarray*}
A(k,\omega>0)&=\sum_l|\langle l|c^{\dagger}_{k\uparrow}|{\rm GS}\rangle|^2\delta(\omega-E_l+E_{\rm GS}),\\
A(k,\omega<0)&=\sum_l|\langle l|c_{k\downarrow}|{\rm GS}\rangle|^2\delta(\omega+E_l-E_{\rm GS}). 
\end{eqnarray*}
Here, $|{\rm GS}\rangle$ and $|l\rangle$ denote the ground state with energy $E_{\rm GS}$ and the excited state with $E_l$, respectively, 
while $c^{\dagger}_{k\sigma}$ creates an electron with momentum $k$ and spin $\sigma$. 
\par
\begin{figure}
\includegraphics[width=8.42cm]{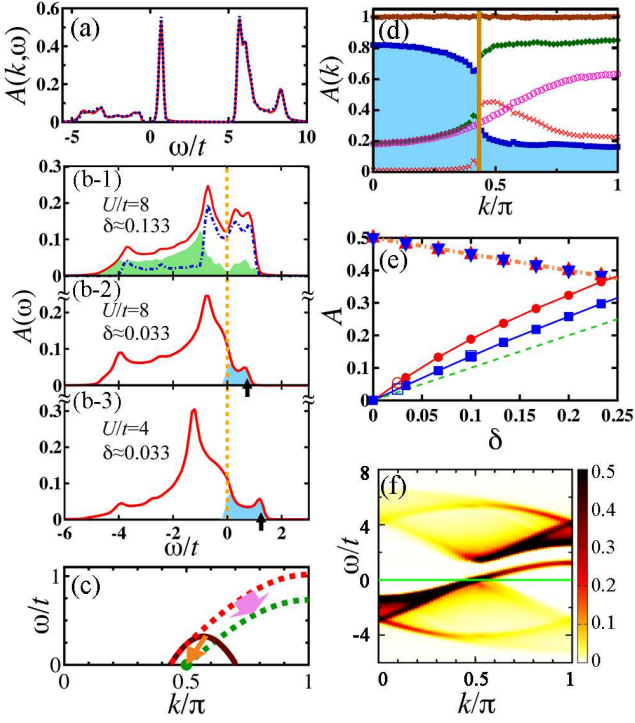}
\caption{(a) $A(k,\omega)$ for $L$=60 (red solid line) and 80 (blue dotted line) at $k$$\simeq$0.8$\pi$ for $U/t$=8 when $\delta$=0.1. 
(b) $A(\omega)$ in the LHB (red lines) for (b-1) $U/t$=8, $\delta$$\simeq$0.133; 
(b-2) $U/t$=8, $\delta$$\simeq$0.033; and (b-3) $U/t$=4, $\delta$$\simeq$0.033. 
In (b-1), the blue dashed-dotted line denotes the contribution of dominant modes, while 
the green region shows the contribution of continua except the dominant modes. 
In (b-2,3), the light blue regions denote the contribution from $\omega$$>$0. 
Arrows indicate $\epsilon$($k$=$\pi$) of Eq. (\ref{eq:spinon}). 
(c) Dispersion relations of the upper edge of the $s$$h^*$ continuum (red dotted line) and the $h^*$ mode with $s$($k_F$) (brown solid line) 
for $U/t$=8 at $\delta$$\simeq$0.133. The green dotted line and the dot at $k$=0.5$\pi$ show those of $\delta$$\rightarrow$0. 
(d) $A(k)$=$\int d\omega A(k,\omega)$ for $U/t$=8 at $\delta$$\simeq$0.133. 
Blue squares with the light blue region denote the contribution from $\omega$$<$0, which is the momentum distribution function in the ground state $n(k)$. 
Green diamonds denote the contribution from $\omega$$>$0; red crosses and pink open circles denote that of the LHB for $\omega$$>$0 
and that of the UHB. The vertical yellow line indicates $k$=$k_F$. 
Brown solid circles denote the total weight at each $k$, which satisfies the sum rule within numerical accuracy. 
(e) Doping dependence of spectral weight $A$. Solid lines show $A$ of the LHB for $\omega$$>$0 
at $U/t$=4 (red circles) and 8 (blue squares). Red and blue triangles denote $A$ for $\omega$$<$0 at $U/t$=4 and 8. 
The green dashed line and orange dashed-dotted line indicate $A$=$\delta$ and (1$-$$\delta$)/2, respectively. 
Open symbols denote data for $L$=80. 
(f) $A(k,\omega)$ for $U/t$=4 at $\delta$$\simeq$0.033.}
\label{fig:Fig1}
\end{figure}
I calculated $A(k,\omega)$ by using the dynamical density-matrix renormalization group (DDMRG) 
method \cite{DDMRG} under the open boundary condition with the number of density-matrix eigenstates $m=120$. 
The data with Lorentzian broadening with half width at half maximum $\eta=0.16t$ were 
deconvolved to those of Gaussian broadening with standard deviation $\sigma=0.1t$. 
The spectral functions 
%$A(k,\omega)$ 
at $k=\pi j/(L+1)$  for $j=1\sim L$ were extrapolated to $k=0$ and $\pi$. 
Figure \ref{fig:Fig1} (a) shows the typical behavior of $A(k,\omega)$. 
Since the difference between the results for $L=60$ and 80 
is small in this scale, this Letter shows the results for $L=60$ unless otherwise mentioned. 
Noting that $A(k,\omega)=A(-k,\omega)=A(k+2\pi,\omega)$, we consider the properties of $0\le k\le\pi$ without loss of generality. 
\par
To identify the dominant modes in $A(k,\omega)$, we use the Bethe ansatz, 
where the wave functions, energies, and momenta of eigenstates are expressed in terms of $\{k_j\}$ 
and $\{\Lambda_\alpha\}$ that satisfy the following Bethe equations \cite{LiebWu}: 
$Lk_j=2\pi I_j+2\sum_{\alpha=1}^M\tan^{-1}\frac{4t(\Lambda_\alpha-\sin k_j)}{U}$ for $j=1\sim N$, and 
$\sum_{j=1}^N\tan^{-1}\frac{4t(\Lambda_\alpha-\sin k_j)}{U}=\pi J_\alpha+\sum_{\beta=1}^M\tan^{-1}\frac{2t(\Lambda_\alpha-\Lambda_\beta)}{U}$ for $\alpha=1\sim M$. 
Here $I_j$ and $J_\alpha$ are integers or half-odd integers. 
We impose the (anti)periodic boundary condition, when $M$ in the ground state is odd (even). 
Because eigenstates are obtained through the Bethe equations once $\{I_j\}$ and $\{J_\alpha\}$ are given, the $\{I_j\}$ and $\{J_\alpha\}$ distributions characterize eigenstates. 
In the ground state, $\{I_j\}$ and $\{J_\alpha\}$ are consecutively distributed around zero. 
Excited states are obtained by creating holes (particles) inside (outside) the consecutive distributions. 
The holes in the $\{J_\alpha\}$ distributions are called spinons, while those in the $\{I_j\}$ distributions are called holons. 
The particles created outside the consecutive $\{I_j\}$ distribution are called antiholons. 
Hereafter, we use the following shorthand notations for the spinon, holon, and antiholon: $s$, $h$, and $h^*$, respectively. 
Noting that the momentum of the eigenstate is given by $K=2\pi(\sum_j I_j+\sum_\alpha J_{\alpha})/L$, 
we define the momenta of $s$, $h$, and $h^*$ as $q_s=2\pi J_s/L$, $q_h=2\pi I_h/L$, and 
$q_{h^*}=2\pi I_{h^*}/L$, respectively, where $J_s$, $I_h$, and $I_{h^*}$ are the positions of $s$, $h$, and $h^*$ in the distributions of $\{J_\alpha\}$ and $\{I_j\}$. 
The momentum ranges are $|q_s|<k_F$, $|q_h|<2k_F$, and $2k_F<|q_{h^*}|<\pi$ with the Fermi momentum $k_F=\pi(1-\delta)/2$.  
\par
\begin{figure}
\includegraphics[width=8.3cm]{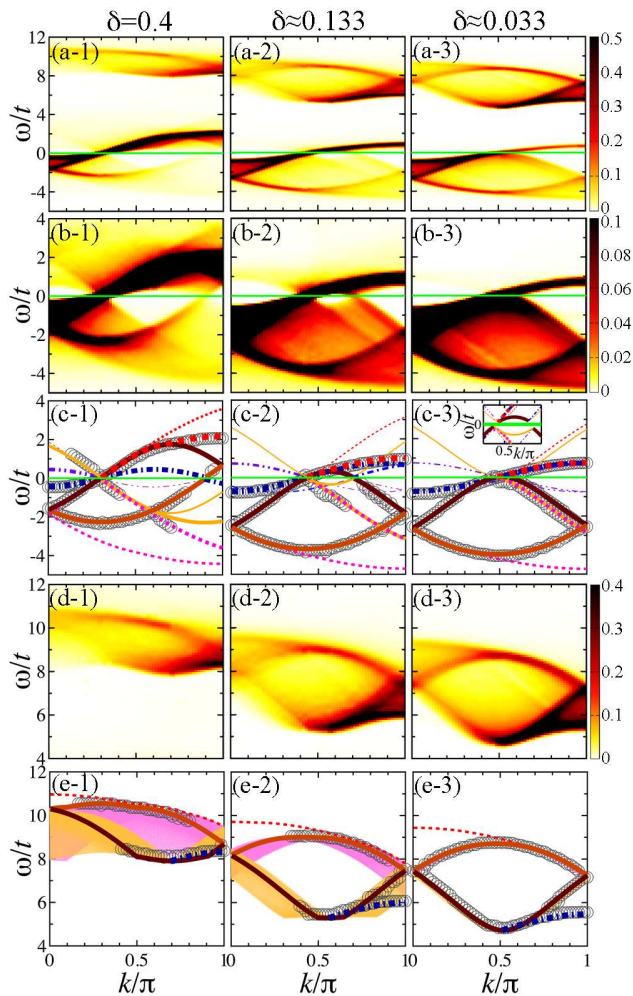}
\caption{(a),(b),(d) $A(k,\omega)$ for $U/t$=8 at $\delta$$\simeq$0.4, 0.133, 0.033 (from the left) 
for (a) overall views, (b) the LHB, and (d) the UHB. 
(c),(e) Dispersion relations obtained using the Bethe ansatz in $L$=240, corresponding to (b),(d). 
Solid lines except those at $\omega$=0 show holon modes and antiholon modes. Blue dashed-dotted lines denote spinon modes. 
Red (pink) dotted lines show upper (lower) edges of continua of $h$$h^*$ with $s$($k_F$) and $h^*$($2k_F$), $s$$h^*$, $h$$h^*$ with $s$($k_F$), 
and $s$$h$$h^*$ (from above) in (c), and $s$$h$$d$ in (e). Purple dashed double-dotted lines in (c) denote spinon modes in 2-$\Lambda$-string solutions. 
Open circles indicate peaks of dominant modes of (b),(d). 
The inset in (c-3) is the close-up near the gapless points. The pink [yellow] region in (e) denotes the $h$$d$ continuum with $s$($-k_F$) [$s$($k_F$)]. 
Spinon [holon] modes in (e) have $h$($\pm2k_F$) and 
%$d$($\pm$($\pi$$-$$2k_F$)) 
$d\left(\pm(\pi-2k_F)\right)$ 
with $q_dq_h$$<$0 
[$s$($\pm k_F$) and $d$($|q_d|$$\simeq$$\pi$$\delta$/2, $q_dq_h$$<$0)].}
\label{fig:Akw}
\end{figure}

{\it Dominant modes for $\omega<0$.$-$} 
Before discussing the Mott transition, we review the dominant modes for $\omega<0$ at $\delta=0.4$ \cite{DDMRGAkw,SchulzSpctra}. 
The dominant mode for $k<k_F=0.3\pi$ near $\omega\lesssim 0$ in Fig. \ref{fig:Akw} (a-1) is identified as 
the $s$ mode without $h$ and $h^*$ [blue dashed-dotted line for $k<k_F$ in Fig. \ref{fig:Akw}(c-1)]. 
The mode slightly below it is identified as the $h$ mode with $s$ having $q_s=k_F$ and $h^*$ having $q_{h^*}=2k_F$. 
Hereafter, we denote it as the $h$ mode with $s$($k_F$) and $h^*$($2k_F$). 
The mode for $k<3k_F=0.9\pi$ is identified as the $h$ mode with $s$($-k_F$) and $h^*$($-2k_F$), 
which is called the shadow band \cite{shadowband}. 
The lower edge of the $s$$h$$h^*$ continuum is indicated by the lowest pink dotted line in Fig. \ref{fig:Akw}(c-1). 
For $\omega\gtrsim 0$, $s$$h^*$ excitations are dominant \cite{SchulzSpctra,Benthien}. 
\par
{\it Pseudogap.$-$} 
As shown by the red line in Fig. \ref{fig:Fig1}(b-1), the momentum-integrated spectral weight [$A(\omega)=\int_{-\pi}^{\pi}\frac{dk}{2\pi}A(k,\omega)$] 
shows reductions near $\omega\simeq0$. This pseudogap behavior originates from the following three properties: 
(i) low-energy property as a Tomonaga-Luttinger liquid \cite{Schulz,AkwQMC}, 
(ii) the band-edge singularity of the 1D dominant modes, and (iii) continua spread above and below $\omega=0$. 
In a Tomonaga-Luttinger liquid, $A(\omega)\propto|\omega|^{\theta}$ for $\omega\rightarrow 0$ with the same exponent as that of the momentum distribution function 
$n(k)-1/2\propto |k-k_F|^{\theta}$ for $k\rightarrow k_F$, where $0<\theta<1/8$ depending on the values of $U/t$ and $\delta$ [Fig. \ref{fig:Fig1}(d)] \cite{Schulz}. 
Because $\theta\rightarrow 1/8$ as $\delta\rightarrow0$, the pseudogap behavior will be significant near the Mott transition. 
Also, the contribution of dominant modes extracted by Gaussian fitting of the peaks in $A(k,\omega)$ 
at each $k$ [blue dashed-dotted line in Fig. \ref{fig:Fig1}(b-1)] shows peaks near the bottom of the $s$ mode for $\omega\lesssim 0$ 
and the top of the $h^*$ mode with $s$($k_F$) [brown solid line near $k\simeq\pi/2$ in Fig. \ref{fig:Akw}(c-2)] for $\omega\gtrsim 0$; 
this implies that the pseudogap behavior can be explained as a dip between the peaks near the band edges of the 1D dominant modes. 
Moreover, the contribution from continua except the dominant modes [green region in Fig. \ref{fig:Fig1}(b-1)] shows reductions 
near $\omega\simeq0$. 
As in Fig. \ref{fig:Akw}(b-2), the continua above and below $\omega=0$ shrink to the gapless points at $k=k_F$ and $2\pi-3k_F$ as $\omega\rightarrow0$. 
The continua for $\omega\gtrsim 0$ are mainly due to the 2-$\Lambda$-string solutions 
where two $\Lambda_{\alpha}$'s are ${\bar \Lambda}\pm\i U/(4t)$ with real ${\bar \Lambda}$ \cite{Takahashi} for $k\lesssim k_F$, 
$s$$h^*$ for $k\simeq2\pi-3k_F$, and $h$$h^*$ excitations with $s$($k_F$) and $h^*$($\pm2k_F$) for $k\simeq2\pi-3k_F$ and $k_F$. 
For $\omega\lesssim 0$, the continua mainly come from the $s$$h$ excitations with $h^*$($\pm2k_F$). 
\par
{\it Hole-pocket behavior.$-$} 
The mode connecting the two gapless points at $k=k_F$ and $2\pi-3k_F$ for $\omega\gtrsim 0$ is identified as the $h^*$ mode with $s$($k_F$). 
On the basis that it has charge character and that the momentum region between these gapless points shrinks as $\Delta k=2\pi\delta$, 
this region can be regarded as a hole pocket. It is expected that the $h^*$ mode will become robust 
as $U/t$ increases because of the reduction in double occupancies. In fact, the $h^*$ mode has been found 
for $U/t\rightarrow\infty$ \cite{shadowband} and in the 1D $t$-$J$ model \cite{antiholon}. 
\par
{\it Spectral-weight transfer.$-$} 
Although the spectral weight $A$ transferred to the LHB for $\omega>0$ equals the amount of doping ($A=\delta$) at $t=0$ \cite{SWtransED}, 
Fig. \ref{fig:Fig1}(e) shows that $A>\delta$ for $t\ne0$. Such behaviors have been discussed in the literature \cite{SWtransED,RMPPhilips,ImadaPRL}. 
[For $\omega<0$, $A=(1-\delta)/2$ as usual.] 
The question here is which mode carries the transferred spectral weight. As in Figs. \ref{fig:Akw}(b-3) and (c-3), 
a considerable weight is carried by the mode of the upper edge of the $s$$h^*$ continuum. 
In the $\delta\rightarrow 0$ limit, where $|q_{h^*}|\rightarrow\pi$, this mode reduces to the $s$ mode with $h^*$($\pm\pi$); 
the dispersion relation $\epsilon(k)$ is obtained as follows: 
\begin{eqnarray}
\label{eq:spinon}
\epsilon&=\frac{2t^2}{U}\int_{-\pi}^{\pi}dq\cos^2q \mbox{ sech}\frac{2\pi t(\Lambda_s-\sin q)}{U},\\
k&=\frac{\pi}{2}+\frac{t}{U}\int_{\Lambda_s}^{\infty}dx\int_{-\pi}^{\pi}dq\mbox{ sech}\frac{2\pi t(x-\sin q)}{U},\nonumber
\end{eqnarray}
by using dressed energies and momenta for $\delta\rightarrow0$ \cite{TakahashiBook,Essler}. 
This gapless mode naturally leads to the 2-spinon continuum at half-filling expressed as 
$E(k)=\epsilon(k_1)+\epsilon(k_2)$ with $k=k_1+k_2-\pi$ \cite{TakahashiBook,Essler}: 
The two gapless modes carrying $S$=1/2 with gapless points at $k=\pm k_F$ cause the gapless $S$=1 spin excitations with the gapless point at $|k|=2k_F$. 
It should be noted that the cosine dispersion relation in the $U/t\rightarrow 0$ limit 
as well as the flat dispersion relation in the $U/t\rightarrow\infty$ limit \cite{Dagotto} are reproduced, 
since $\epsilon(k)\simeq-2t\cos k$ for $U\ll t$ and $\epsilon(k)\simeq-\frac{\pi J}{2}\cos k$ 
with $J\equiv 4t^2/U$ for $U\gg t$ \cite{TakahashiBook} for $|k\pm\pi|<\pi/2$. 
This implies that the spectral weights transferred by infinitesimal doping spread up to the energy of $O(t)$ for small $U/t$ and $O(J)$ 
for large $U/t$, as in Figs. \ref{fig:Fig1}(b-2)$-$(c), (f), and \ref{fig:Akw}(b-3). 
This behavior contrasts with band-insulator-to-metal transitions, 
where the transferred spectral weight remains within the chemical potential shift $\Delta\mu$ of $O(\delta^{2/D})$ for $\delta\rightarrow0$ 
in $D$ dimensions. 
\par
{\it Upper Hubbard band.$-$} 
To identify the dominant modes in the UHB, we consider the solutions with a $k$-$\Lambda$ string, 
i.e., a pair of complex $k_j$'s with a nonzero imaginary part, which represents a pair of electrons \cite{Takahashi}. 
Since such solutions have energies of $O(U)$ in the large $U/t$ regime \cite{Takahashi,TakahashiBook}, 
it is natural to expect that they are relevant for the UHB \cite{Woynarovich,Benthien}. 
In fact, it has been found that solutions with a string length greater than 1 have considerable spectral weights 
in the high-energy regime for quasi-1D spin models \cite{1DH, q1DH} and spinless-fermion chains \cite{1DSF,1DSF_DMRG}. 
The $k$-$\Lambda$ string is characterized by a half-odd integer or an integer ${\bar I}_1$. 
We call the quasiparticle (QP) for the $k$-$\Lambda$ string a doublon and denote it as $d$; 
the momentum is defined as $q_d=2\pi{\bar I}_1/L$ in $|q_d|<\pi-2k_F=\pi\delta$ \cite{Takahashi}. 
We define the sign of $q_d$ such that $k=-q_s-q_h+q_{h^*}+q_d$. 
\par
At half-filling, $A(k,\omega)$ of the UHB is symmetric with that of the LHB because of the particle-hole symmetry. 
In the small-$\delta$ regime, $A(k,\omega)$ remains almost symmetric, as in Figs. \ref{fig:Fig1}(f) and \ref{fig:Akw}(a-3). 
In the $\delta\rightarrow0$ limit, where $|q_d|\rightarrow0$, the dominant modes in the UHB are 
characterized by $s$ and $h$ in the $k$-$\Lambda$-string solutions and have essentially the same character as that in the LHB. 
Namely, the low-energy mode for $k\gtrsim k_F\simeq\pi/2$ in Fig. \ref{fig:Akw}(d-3) is identified 
as the $s$ mode with $h$($\pm 2k_F$) and $d$($q_d\simeq0$), as in Fig. \ref{fig:Akw}(e-3). 
The mode slightly above it and the low-energy mode for $k\lesssim k_F$ are both identified as the $h$ mode with $s$($k_F$) and $d$($q_d\simeq0$). 
The high-energy mode is identified as the $h$ mode with $s$($-k_F$) and $d$($q_d\simeq0$), which corresponds to the shadow band. 
For larger doping, the dominant modes with charge character in Figs. \ref{fig:Akw}(d-1,2) 
remain within the $h$$d$ continua with $s(\pm k_F)$ [yellow and pink regions in Figs. \ref{fig:Akw}(e-1,2)] 
and appear to have $d$($|q_d|\simeq\pi\delta/2$, $q_dq_h<0$) [solid lines in Figs. \ref{fig:Akw}(e-1,2)]. 
\par
{\it Discussion.$-$} 
To obtain an intuitive understanding of Mott transitions, two scenarios have been considered. 
One is that the effective carrier density ($n^*$) vanishes, as in the case of metal-to-band-insulator transitions. 
The other is the effective-mass ($m^*$) divergence in the Fermi liquid theory. 
The present results show that spectral weights are mainly transferred to the $h^*$ mode with $s$($k_F$) and 
the mode of the upper edge of the $s$$h^*$ continuum [Figs. \ref{fig:Fig1}(b-1)$-$(c)]. 
The former behaves like $n^*\rightarrow 0$, because the hole pocket shrinks as $\delta\rightarrow 0$. 
The latter may seem to behave as $m^*\rightarrow\infty$, because its spectral weight decreases [Fig. \ref{fig:Fig1}(e)]. However, in a Fermi liquid, 
the renormalized bandwidth near the chemical potential shrinks as $\epsilon(k)\propto1/m^*$ for $m^*\rightarrow\infty$, 
which is not the case for the present results wherein $\epsilon(k)$ in Eq. (\ref{eq:spinon}) remains $O(t)$ or $O(J)$ [Fig. \ref{fig:Fig1}(c)]. 
Thus, a simple classification of $n^*\rightarrow 0$ or $m^*\rightarrow\infty$ is not accurate for the Mott transition of the Hubbard chain; 
the Mott transition is rather characterized as a loss of charge character from the mode having both spin and charge characters, 
while the dispersion relation of the spin part remains gapless and dispersing. 
\par
The nature of the Mott transition in the Hubbard chain may be more clearly understood through comparisons 
with the results of a recent cellular dynamical mean-field study on the Mott transition of the 2D Hubbard model in Ref.~\cite{ImadaPRL}, 
where the following features have been suggested. 
(i) The phase transition from a large Fermi surface to hole pockets occurs at nonzero $\delta$. 
(ii) The transferred spectral weight behaves as $A=\delta+\langle n_{i\uparrow}n_{i\downarrow}\rangle$ after tiny doping. 
(iii) A gap fully opens between the band forming the hole pockets and that of ingap states for $\omega>0$ below the UHB, leading to the zero surface of the Green function \cite{Kotliar}. 
(iv) The pseudogap can be attributed to doublon (doubly occupied site)-holon (empty site) binding \cite{RMPPhilips} relaxed due to doping. 
\par
In contrast, for the Hubbard chain, (i) there is no phase transition when $\delta\ne 0$: The gapless point at $k=2\pi-3k_F$ persists 
even in the large doping regime, where the gapless point goes beyond the zone boundary ($k=\pi$) and is located at $k=3k_F$, as in Fig. \ref{fig:Akw}(b-1). 
(ii) The transferred spectral weight does not follow $A=\delta+\langle n_{i\uparrow}n_{i\downarrow}\rangle$ 
and gradually reaches zero as $\delta\rightarrow0$, as in Fig. \ref{fig:Fig1}(e). 
(iii) The cosinelike mode and the antiholon mode are both gapless at $k=k_F$: There is no gap between them at $k=k_F$. 
The gapless cosinelike mode is related to the gapless spin excitations. 
(iv) Double occupancy in the LHB does not behave as a QP. 
The QP for a pair of electrons is defined in the $k$-$\Lambda$-string solutions, which are relevant only for the UHB: 
The pseudogap behavior in the LHB is not related to the QP. 
Understanding and explaining the above differences can be an interesting topic for future study. 
\par
{\it Summary.$-$} 
The single-particle spectral properties near the Mott transition were investigated by using the Bethe ansatz and DDMRG method in the 1D Hubbard model. 
Characteristic spectral features near the Mott transition, such as the pseudogap, hole-pocket behavior, 
spectral-weight transfer, and upper Hubbard band, were explained in a unified manner in terms of spinons, holons, antiholons, and doublons. 
A remarkable feature is the emergence of the gapless mode extending up to $O(t)$ [$O(J)$] for small [large] $U/t$ by infinitesimal doping. 
This mode is related to the spin excitations at half-filling and cannot be interpreted by either the rigid-band picture or the Fermi liquid theory. 
The present results have similarities to the anomalous spectral features observed in high-$T_c$ cuprates \cite{kink,holepocket} 
and some aspects of the 2D Hubbard model \cite{RMPPhilips,ImadaPRL,Kotliar}. 
I expect that generic features of Mott transitions in finite dimensions can be deduced from the present results. 
\par
I am grateful to S. Fujimoto and M. Arikawa for helpful discussions. This work was supported by 
WPI Initiative on Materials Nanoarchitectonics, 
MEXT, Japan, and KAKENHI 22014015. 
%\bibliography{apssamp}

\end{document}